\documentclass[twocolumn,pra,showpacs]{revtex4}

\usepackage[dvips]{graphicx}

\begin{document}


\title{Parity nonconservation in Atomic Zeeman Transitions}

\author{E.J. Angstmann}
\affiliation{School of Physics, University of New South Wales,
Sydney 2052, Australia}
\author{T.H. Dinh}
\affiliation{School of Physics, University of New South Wales,
Sydney 2052, Australia}
\author{V.V. Flambaum}
\affiliation{School of Physics, University of New South Wales,
Sydney 2052, Australia}

\date{\today}

\begin{abstract}
  We discuss the possibility of measuring nuclear anapole moments in atomic Zeeman
transitions and perform the necessary calculations. Advantages of using Zeeman transitions
include variable transition frequencies and the possibility of enhancement of parity nonconservation
effects.
\end{abstract}

\maketitle

  Experiments on parity nonconservation in atoms have measured the
 electron-nucleon weak interaction and have provided very accurate
tests of the Standard Model  at low energies (see, e.g., review \cite{GF}).
It has been pointed out in Ref. \cite{Flambaum} that atomic experiments
can be used to measure the nuclear anapole moment and study parity nonconserving (PNC)
nuclear forces. The measurement of the anapole moment of $^{133}$Cs nucleus was
reported in Ref. \cite{Wieman}. However, the strength of the PNC nuclear forces
extracted from this measurement seem to disagree with the limits
on these forces extracted from the $^{205}$Tl anapole measurement
performed in another atomic experiment \cite{Fortson}. Moreover, the general
situation with PNC nuclear forces at the moment is controversial since
different experiments give contradictory results (see, e.g., review \cite{GF}).
This situation requires new measurements of nuclear anapole moments.

The PNC interaction in atoms can be split into a nuclear spin-dependent (NSD) and nuclear spin-independent (NSI)
 part. The NSI part is due to the nuclear weak charge, and the PNC effects produced by the NSI part are
two orders of magnitude larger than the PNC effects produced by the NSD part.
 There are three contributions to the NSD part of the PNC interaction:
 the nuclear anapole moment, the NSD part of the electron-nucleus interaction,
 and the combination of the NSI electron-nucleus weak interaction and the hyperfine interaction.
 For heavy atoms the anapole moment contribution dominates since it grows as $A^{2/3}$\cite{Flambaum,FSK,FK}.
In experiments \cite{Wieman,Fortson} the NSD contribution was separated from the much larger NSI contribution. The NSD contribution is different in different hyperfine components of an optical transition and can be
 extracted from the difference of the PNC effects in these components. An alternative method to measure the
NSD interaction was suggested in Ref. \cite{Khriplovich_1975}.
 The NSI part is zero in transitions between hyperfine terms of a ground state electronic level.
The measurements of PNC in the hyperfine transitions is sensitive to the NSD PNC interaction only.
Unfortunately, no such experiment has been performed.

It was suggested in Refs. \cite{Zeeman} (see also review
\cite{Budker} and references therein) that the transitions between
the Zeeman components of the hyperfine levels of a ground state atom
in a magnetic field may be more convenient for the measurements of
PNC induced by the NSD interaction, than transitions between the
hyperfine levels themselves. The NSI interaction still does not
contribute here, therefore, the measurement of the PNC effects in
Zeeman transitions can provide a value of the nuclear anapole
moment. One advantage of using these transitions is that for weak
external magnetic fields,
 the transition frequency is proportional to the field and so can be chosen arbitrarily small,
and determined from the conditions required for optimal experimental setup.
 Advantages of the low frequency
 were explained in the proposal of possible experiments in hydrogen, potassium and cesium hyperfine transitions \cite{Gorshkov}. At low frequencies one does not need a resonator
to produce the electromagnetic field. This
greatly simplifies the experiment and allows the frequency to be set to the same value for different atoms.
 It should be noted that the PNC E1 amplitudes of the Zeeman transitions in weak magnetic fields
 are suppressed by a factor $\hbar\omega/\Delta E_{hf}$, where $\omega$ is the transition frequency
 and $\Delta E_{hf}$ is the hyperfine splitting. However, the E1 amplitude increases with nuclear
 charge as $d \propto Z^2A^{2/3}R_a$, where $R_a$ is a relativistic correction factor (see, e.g.
 \cite{Khriplovich,GF} ), while the hyperfine splitting increases as $\Delta E_{hf} \propto Z$.
 Combining these results we can see that the E1 amplitude of Zeeman transitions increases faster
 than $ZA^{2/3}$ for heavy atoms in weak magnetic fields at a fixed transition
 frequency. This suggests that heavy atoms should be considered as
 candidates for an experiment.

  Another possible advantage appears for Zeeman transitions in a strong magnetic
field \cite{Zeeman,Gorshkov}. Parity nonconservation
appear due to interference
 between the PNC electric dipole amplitude, $d$, and the magnetic dipole amplitude,
 $\mu$,
$P \sim Im[2d\mu/(|\mu|^2+|d|^2)] \approx 2Im(d/\mu)$.
 In a strong magnetic field an eigenstate is approximately a product
 of the nuclear spin state
times the electron wave function. The $M1$ amplitude, $\mu$, between the Zeeman
 states corresponding to nuclear spin-flip is very small, since the nuclear magnetic moment is $\sim$1000 times smaller than the electron
magnetic moment (in reality the contribution from the small admixture of the state with a different electron angular momentum projection will always dominate $\mu$). The magnitude of the PNC amplitude, $d$, is not suppressed for such transitions.
 This enhances the value of PNC effects  ($\sim d/\mu$) containing $\mu$ in the denominator.


In this paper we will consider three different experimental schemes:
no magnetic field, a weak magnetic field ($g\mu_B B \ll \Delta
E_{hf}$), and a strong magnetic field ($g\mu_B B \gg \Delta E_{hf}
\gg g_I \mu_N B$). For each we shall calculate the ratio
$R=d/\omega\mu$, where $d$ is the PNC E1 amplitude, $\mu$ is the M1 amplitude
 and $\omega$ is the transition frequency. A
promising transition will have a relatively large $R$, since the PNC
E1 amplitude, $d$, will be large, the background parity conserving
M1 amplitude, $\mu$, will be relatively small, and ideally the
transition frequency should also be low, as mentioned previously.

We consider atoms and ions with the electron angular momentum  $J=1/2$.
In this case the only allowed values for the total atomic angular momentum
are $F=I\pm 1/2$, where $I$ is the nuclear spin. In the presence of a magnetic
field, states with a fixed $F$ are split into states with a fixed projection,
$M$, onto the external field direction. The total angular momentum is no longer
conserved, there is a mixing of states with the same projection, $M$,
resulting in the appearance of the new states:
\begin{eqnarray}
|\Psi^{(1)},M\rangle & = & c_{1}^{(1)}|I+1/2,M\rangle + c_2^{(1)}|I-1/2,M\rangle \\
|\Psi^{(2)},M\rangle & = & c_{1}^{(2)}|I+1/2,M\rangle +
c_2^{(2)}|I-1/2,M\rangle.
\end{eqnarray}
The mixing coefficients, $c$, in these equations are calculated from the secular
equations  with the hyperfine interaction and the interaction with the external
magnetic field taken together. In the low field limit
\begin{eqnarray*}
|\Psi^{(1)},M\rangle & \to & |I+1/2,M\rangle \\
|\Psi^{(2)},M\rangle & \to & -|I-1/2,M\rangle.
\end{eqnarray*}
In the high field limit the coupling between $I$ and $J$ is broken and it is no
longer proper to consider the states in terms of $F$, instead we need to consider
them in terms of $m_I$ and $m_J$. In this limit we have the states
\begin{eqnarray*}
|\Psi^{(1)},M\rangle & \to & |m_{I}=M-1/2,m_{J}=1/2\rangle \\
|\Psi^{(2)},M\rangle & \to & |m_{I}=M+1/2,m_{J}=-1/2\rangle.
\end{eqnarray*}
It makes no sense to divide the terms into hyperfine and Zeeman
transitions in the high field limit because $F$ is not conserved,
nevertheless we shall call the $|\Psi^{(1)}\rangle
\rightarrow|\Psi^{(2)}\rangle$ transitions the hyperfine ones and
the $|\Psi^{(1)},M\rangle\rightarrow|\Psi^{(1)},M\pm 1\rangle$ , and
$|\Psi^{(2)},M\rangle\rightarrow|\Psi^{(2)},M\pm 1\rangle$ the
Zeeman ones.

\begin{figure}[htb]
\begin{center}
\includegraphics[width=0.6\textwidth]{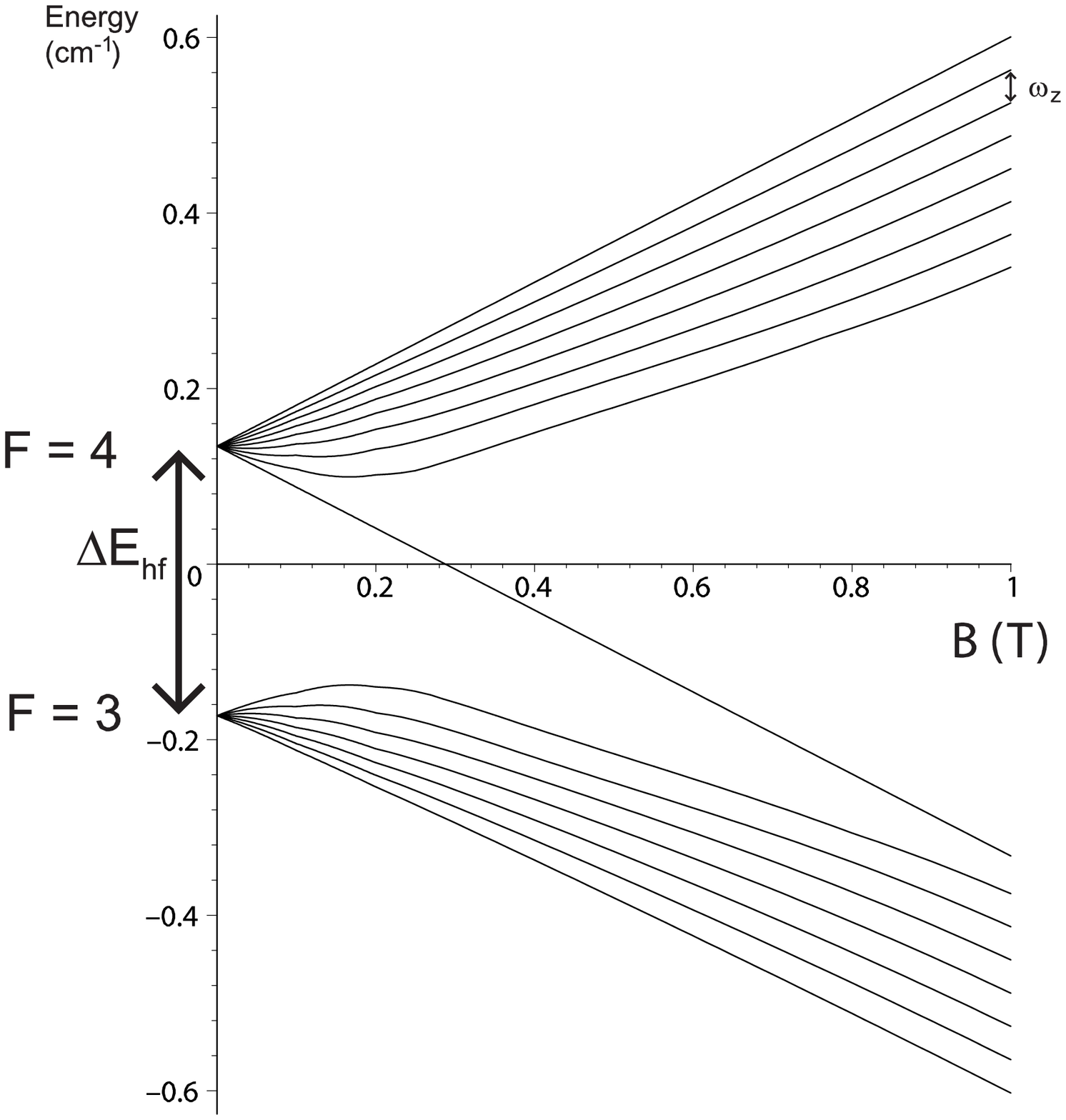}
\caption{\label{fig:Cs} Zeeman splitting of the hyperfine terms of $^{133}$Cs
as a function of magnetic field. In weak fields the splitting between hyperfine
levels (marked $\Delta E_{hf}$) is fairly constant and the splitting between
Zeeman levels increases linearly with the field. In strong fields the splitting
between the hyperfine levels increases linearly with the field while the splitting
between the Zeeman levels reaches a constant separation (marked $\omega_{Z}$).}
\end{center}
\end{figure}

The dependence of the $^{133}$Cs $6s_{1/2}$ energy levels on the
external magnetic field is presented in Figure \ref{fig:Cs}. In weak
magnetic fields the energy difference of the hyperfine term
$|\Psi^{(1)}\rangle$ and $|\Psi^{(2)}\rangle$ is constant and equal
to $\Delta E_{hf}$ while the energy difference of the Zeeman terms
is proportional to the field. In the strong field the hyperfine
transition energy increases in proportion to the field while the
Zeeman transition energy is constant and equal to $\Delta
E_{hf}/(2I+1)$. Here and below we neglect small terms $\sim \mu_n B$
where $\mu_n$ is the nuclear magnetic moment.

For the hyperfine transitions without the external field the ratio $R$ is given by
\begin{eqnarray}
R_{0} & = &  \frac{\langle\Psi^{(2)},M|d|\Psi^{(1)},M\rangle}{\omega\langle\Psi^{(2)},
M|\mu|\Psi^{(1)},M\rangle} \nonumber\\
 & = &  \Big(\frac{I+1/2}{I(I+1)}\Big)^{1/2}\frac{\langle I-1/2||d||I+1/2\rangle}{g\mu_{B}\Delta E_{hf}}.
\end{eqnarray}
Here $g$ is the atomic magnetic (Land\'{e}) $g$-factor: $g=2$ for
the $s_{1/2}$ valence electron (K, Rb, Cs, Ba$^+$, Au, Fr), and
$g=2/3$ for the $p_{1/2}$ valence electron (Tl).

In weak fields the E1 amplitudes behave just like the transition
frequencies (they are proportional to the field for Zeeman
transitions and constant for hyperfine ones), while the M1
amplitudes are independent of the weak field. The $R$ parameters,
given by
\begin{displaymath}
R=\frac{\langle\Psi^{(1)},M|d_+|\Psi^{(1)},M+1\rangle}{\omega\langle\Psi^{(1)},M|\mu_+|\Psi^{(1)},M+1\rangle},
\end{displaymath}
for hyperfine transitions are independent of the field while it is
weak. For Zeeman transitions they are given by
\begin{eqnarray}
R_{|\Psi^{(1)},M\rangle\rightarrow|\Psi^{(1)},M\pm1\rangle}=2 IR_{0},\\
R_{|\Psi^{(2)},M\rangle\rightarrow|\Psi^{(2)},M\pm1\rangle}=2
(I+1)R_{0}.
\end{eqnarray}
Here we use spherical components of the electric and magnetic dipole
operators defined as $d_+=(d_x+id_y)/\sqrt{2}$ and
 $\mu_+=(\mu_x+i\mu_y)/\sqrt{2}$.

In strong magnetic fields the transitions $|\Psi^{(1)},M\rangle
\rightarrow |\Psi^{(2)},M\pm 1\rangle$ are not interesting since the
transition $|\Psi^{(1)},M\rangle \rightarrow |\Psi^{(2)},M+1\rangle$
corresponds to $\Delta I_{z}=2$, resulting in the E1 amplitude being
forbidden due to selection rules, while for the transition
$|\Psi^{(1)},M\rangle \rightarrow |\Psi^{(2)},M-1\rangle$ both the
M1 and E1 amplitude are independent of the magnetic field and so the
magnetic field dependence of $R$ comes entirely from $\omega$ which
increases linearly with magnetic field, resulting in $R$ decreasing
inversely to the magnetic field. By contrast, in the transition
$|\Psi^{(1)},M\rangle \rightarrow |\Psi^{(2)},M\rangle$ the E1
amplitude is independent of the external magnetic field. The
parameter $R_{|\Psi^{(1)},M\rangle \rightarrow
|\Psi^{(2)},M\rangle}$ can also be shown to be independent of the
field and always equal to $R_0$. The E1 amplitude of the Zeeman
transitions in a strong magnetic field reach a constant value,
comparable to the E1 amplitude of the hyperfine transitions. The
frequencies also reach a constant value, given by $\omega = \Delta
E_{hf}/(2I+1)$, while the M1 transition amplitudes decrease in
proportion to the field. This results in the R parameters increasing
with the field
\begin{eqnarray}
R_{|\Psi^{(1)},M\rangle\rightarrow|\Psi^{(1)},M\pm1\rangle} & = &
R_{|\Psi^{(2)},M\rangle\rightarrow|\Psi^{(2)},M\pm1\rangle} \nonumber\\
& = & 2(I+1/2)R_{0}x,
\end{eqnarray}
where $x=g\mu_BB/\Delta E_{hf}$.

\begin{figure}[h]
\begin{center}
\includegraphics[width=0.6\textwidth]{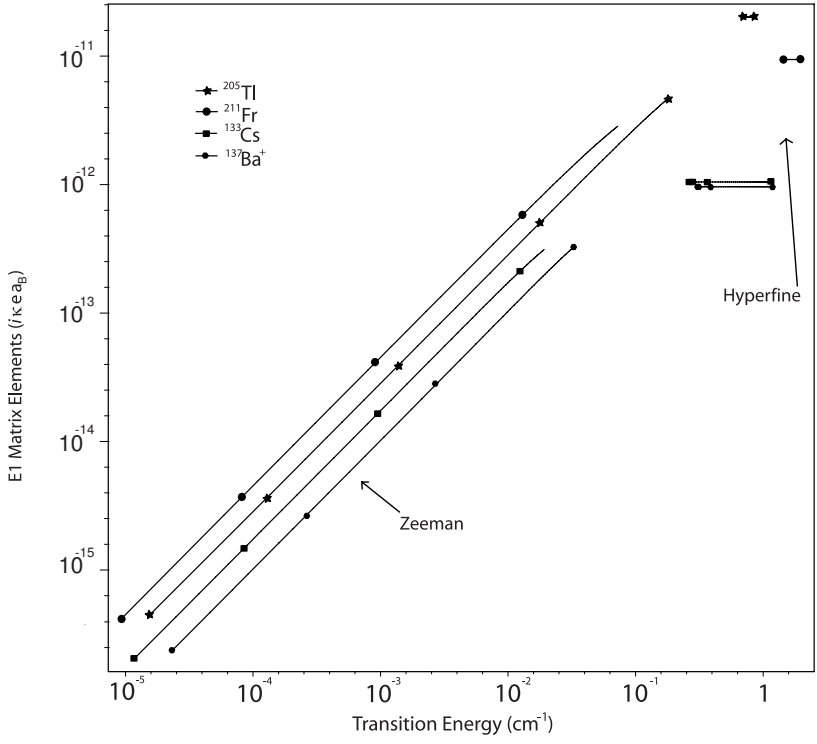}
\caption{\label{fig:d} P-odd E1 transition amplitudes between Zeeman and
hyperfine terms of different atoms as a function of transition frequency
and external magnetic field. The points represent the elements and also
represent the strength of the magnetic field, from left to right 0.0001
T, 0.001 T, 0.01 T, 0.1 T and 1T. The graph shows that the E1 amplitude
of Zeeman transitions increases linearly with the field for small fields
 before reaching a constant value, but retains a constant value for
 hyperfine transitions, independent of the field. It should be noted that
 in the case of Zeeman transitions, the transition energy and E1 amplitude
 reach a constant maximum value at a critical field given by $B = \frac{\Delta E_{hf}}{g \mu_{B}}$
 ($x = 1$), in this graph the lines for the Zeeman transitions end at half this value ($x = 1/2$),
 which is less than 1T for several atoms. The hyperfine transition energy is almost constant for
 small fields (as can be seen in Fig. \ref{fig:Cs}), resulting in several points appearing at the same position. }
\end{center}
\end{figure}

\begin{figure}[h]
\begin{center}
\includegraphics[width=0.6\textwidth]{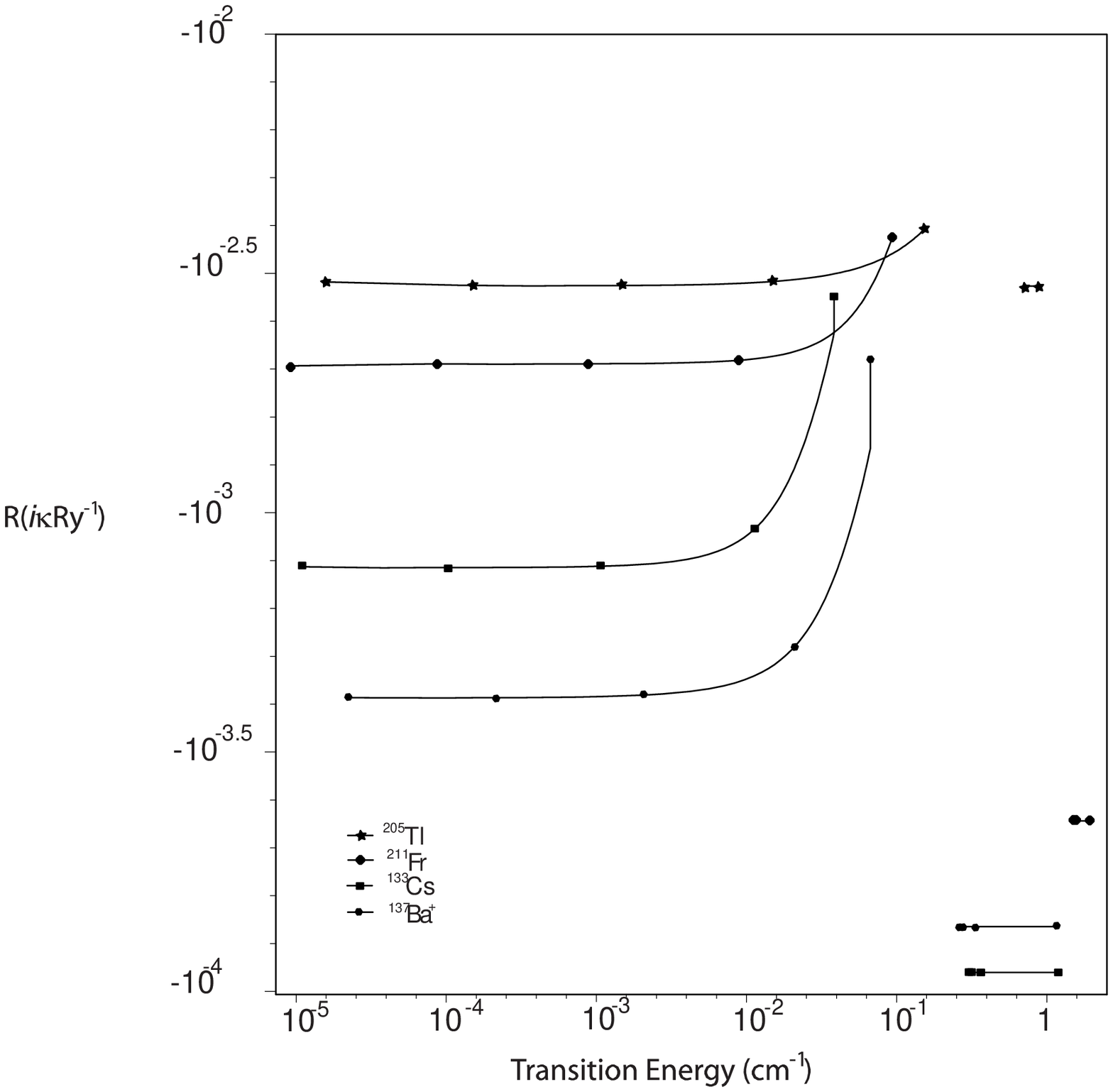}
\caption{\label{fig:R} The ratio, $R$, of the PNC E1 amplitude, $d$, to the background
 ($\omega\mu$) as a function of external field and transition frequency.
 The external magnetic field is once again represented by the points, ranging from
 0.0001 T up to 1 T from, left to right. This graphs shows that $R$ for Zeeman
 transitions can be significantly higher than for hyperfine transitions in the same atom.}
\end{center}
\end{figure}

Calculated $R$ parameters for a collection of atoms are presented in
Table \ref{tab:values}. In Tables \ref{tab:omega}, \ref{tab:E1} and
\ref{tab:M1} we present $\omega$, $d$ and $\mu$ for the different
magnetic fields. The dimensionless coupling constant, $\kappa$,
consisting mainly of the anapole moment of the nucleus has been left
as a free parameter, we present an estimate of $\kappa_{a}$ in Table
\ref{tab:E1} to give an approximation of the size of this term. The
P-odd E1 transition amplitudes, $d$, $R$ parameter and transition
frequencies, $\omega$ all depend upon the magnetic field, but it is
useful to consider $d$ and $R$ as a function of $\omega$. These
relationships are presented in Figures \ref{fig:d} and \ref{fig:R}.
Moving from left to right in these graphs corresponds to an
increasing magnetic field. These figures show that for transitions
with the same frequency, Zeeman transitions have a larger P-odd E1
amplitude as well as a larger ratio of the E1 amplitude to the
P-even M1 amplitude than the hyperfine transitions.

We conclude that Zeeman transitions in heavy atoms have some advantages over the
hyperfine transitions that have been considered previously. The biggest advantage
is that the transition frequencies can be
 tuned to make  an experiment  easier to perform, increase sensitivity and provide additional control
 of systematic effects  by varying the magnetic field.
 Zeeman transitions should be considered as candidates for experiments to measure the anapole moment of nuclei.

\begin{table*}
\caption{We present the ratio, $R$, for a range of atoms and ions with J=1/2. $R$ is presented for no magnetic field, a small magnetic field and a large magnetic field, $x=\frac{g\mu_{B}B}{\Delta E_{hf}}$ is a dimensionless parameter representing the magnetic field strength. The values of the PNC reduced matrix elements, $\langle F_{F}||d||F_{I}\rangle$, are taken from \cite{Safronova}. Note that
we use the convention for $\kappa$ and the PNC reduced dipole matrix
elements used in the earlier work \cite{Flambaum}. To convert to the
notation used in \cite{Safronova} $\kappa$ must be multiplied by the
factor $(-1)^{I+1/2+l}\frac{(I+1/2)}{I(I+1)}$ where $l$ is the
orbital angular momentum of the unpaired nucleon. }
\begin{tabular}{c c c c c c c c c}\label{tab:values}
Atom/ion & I & $nl$ & $F_{I}-F_{F}$ & $\Delta E_{hf}$ & $\langle
F_{F}||d||F_{I}\rangle$ &
\multicolumn{3}{c}{R ($ 10^{-4}i\kappa Ry^{-1}$)} \\
 & & & & $cm^{-1}$ & $10^{-13}i\kappa e a_{B}$& $|\Psi^{(1)},M\rangle\rightarrow|\Psi^{(2)},
 M\rangle$ & \multicolumn{2}{c}{$|\Psi^{(1)},M \rangle \rightarrow|\Psi^{(1)},M\pm1\rangle$}\\
 & & & & & & $x=0$ & $x \ll 1$ & $x \gg 1$\\\hline
$^{39}$K & 1.5 & $4s$ & 1-2 & 0.0154\footnotemark[1]& -1.18 & -0.85 & -2.5 & -3.4$x$\\
$^{85}$Rb & 2.5 & $5s$ & 2-3 & 0.101\footnotemark[2] & -8.74 & -0.76 & -3.8 & -4.6$x$\\
$^{87}$Rb & 1.5 & $5s$ & 1-2 & 0.228\footnotemark[3] & 7.27 & 0.35 & 1.1 & 1.4$x$ \\
$^{133}$Cs & 3.5 & $6s$ & 3-4 & 0.307\footnotemark[1] & -43.8 & -1.1 & -7.6 & -8.6$x$\\
$^{137}$Ba$^{+}$ & 1.5 & $6s$ & 1-2 & 0.268\footnotemark[4] & -32.9 & -1.3 & -4.0 & -5.4$x$\\
$^{197}$Au & 1.5 & $6s$ & 1-2 & 0.203\footnotemark[5] & 85.4 & 4.6 & 14. & 18.$x$\\
$^{205}$Tl & 0.5 & $6p_{1/2}$ & 0-1 & 0.71\footnotemark[6] & -400\footnotemark[8] & -29. & -29. & -59.$x$\\
$^{211}$Fr & 4.5 & $7s$ & 4-5 & 1.451\footnotemark[7] & -481\footnotemark[9] & -2.2 & -20. & -22.$x$\\
\end{tabular}
\footnotetext[1]{Reference \cite{Happer}}
\footnotetext[2]{Reference \cite{Vanier}}
\footnotetext[3]{Reference \cite{Bize}}
\footnotetext[4]{Reference \cite{Blatt}}
\footnotetext[5]{Reference \cite{Dahman}}
\footnotetext[6]{Reference \cite{Lurio}}
\footnotetext[7]{Reference \cite{Lieberman}}
\footnotetext[8]{A significantly different value of -141 was obtained for this reduced matrix element in \cite{Kozlov1}. }
\footnotetext[9]{A similar value of -445 was calculated in \cite{Kozlov2}.}
\end{table*}

\begin{table}
\caption{We present the transition energy (in units $cm^{-1}$) for
the atoms and transitions that have been considered in this paper, $x=\frac{g \mu_{B}B}{\Delta E_{hf}}$ is a dimensionless parameter related to the magnetic field strength.
The $B=0$ column contains the hyperfine splitting which is
referenced in Table \ref{tab:values}, the next two columns contain
the Zeeman splitting between adjacent levels, in cesium these
energies correspond to the energy difference marked $\omega_{Z}$ on
Fig. \ref{fig:Cs}.}
\begin{tabular}{c c c c }\label{tab:omega}
Atom/ion & B=0 & $g\mu_B B \ll \Delta E_{hf}$ & $g\mu_B B \gg \Delta E_{hf} \gg g_I \mu_N B$ \\
\hline
theory & $\Delta E_{hf}$ & $\frac{x \Delta E_{hf}}{2I+1}$ & $\frac{\Delta E_{hf}}{2I+1}$ \\
$^{39}$K & 0.0154 & 0.0039$x$ & 0.0039 \\
$^{85}$Rb & 0.101 & 0.0168$x$ & 0.0168 \\
$^{87}$Rb & 0.228 & 0.0570$x$ & 0.0570  \\
$^{133}$Cs & 0.307 & 0.0384$x$ & 0.0384 \\
$^{137}$Ba$^{+}$ & 0.268 & 0.0670$x$ & 0.0670 \\
$^{197}$Au & 0.203 & 0.0508$x$ & 0.0508 \\
$^{205}$Tl & 0.71 & 0.355$x$ & 0.355 \\
$^{211}$Fr & 1.451 & 0.1451$x$ & 0.1451 \\
\end{tabular}
\end{table}

\begin{table*}
\caption{Here we present the E1 amplitudes as a function of field
in units of $i\kappa ea_{B}$. We include a rough estimate of
$\kappa_a$. For each atom we present the transition with the largest E1 amplitude in a low field.}
\begin{tabular}{c c c c c c }\label{tab:E1}
Atom/ion & $\kappa_a$\footnotemark[1] & $\langle\Psi^{(1)},M|d_{z}|\Psi^{(2)},M\rangle$\footnotemark[2]
& transition\footnotemark[3] & $\langle\Psi,M+1|d_{+}|\Psi,M\rangle$\footnotemark[4] & $\langle\Psi
,M+1|d_{+}|\Psi,M\rangle$\footnotemark[5] \\
 & & $B = 0$ & & $g\mu_B B \ll \Delta E_{hf}$ & $g\mu_B B \gg \Delta E_{hf} \gg g_I \mu_N B$ \\
\hline
$^{39}$K & 0.17 & $3.7 \times 10^{-14}$ & I & $2.3 \times 10^{-14} x$ & $ 2.7 \times 10^{-14}$  \\
$^{85}$Rb & 0.28 & $2.4 \times 10^{-13}$ & I & $1.6 \times 10^{-13} x$ & $ 1.7 \times 10^{-13}$  \\
$^{87}$Rb & 0.28 & $-2.3 \times 10^{-13}$ & I & $-1.4 \times 10^{-13} x$ & $-1.6 \times 10^{-13}$ \\
$^{133}$Cs & 0.38 & $1.1 \times 10^{-12}$ & I & $7.2 \times 10^{-13}x$ & $7.6 \times 10^{-13}$ \\
$^{137}$Ba$^{+}$ & 0.38 & $1.0 \times 10^{-12}$ & I & $6.4 \times 10^{-13} x$ & $7.4 \times 10^{-13}$ \\
$^{197}$Au & 0.49 & $-2.7 \times 10^{-12}$ & I & $-1.7 \times 10^{-12} x$ & $-1.9 \times 10^{-12}$ \\
$^{205}$Tl & 0.50 & $2.3 \times 10^{-11}$ & II & $1.2 \times 10^{-11} x$ & $1.6 \times 10^{-11}$ \\
$^{211}$Fr & 0.51 & $1.1 \times 10^{-11}$ & II & $7.5 \times 10^{-12} x$ & $7.7 \times 10^{-12}$ \\
\end{tabular}
\footnotetext[1]{These are calculated using the formula
$\frac{9}{10}\frac{\alpha \mu} {m_p r_0}A^{2/3}g_p$, where $\alpha$
is the fine structure constant, $\mu$ is the magnetic moment in
nuclear magnetons of the external nucleon (a proton for all the
cases above), $m_p$ is the mass of the external nucleon (ie proton
mass), $r_0= 1.2$ fm, and $g_p= 4.5$ is the strength of the weak nuclear potential in units of the Fermi constant $G_{F}$. More details about how anapole
moments can be calculated can be found in the review \cite{GF}.}
\footnotetext[2]{These are calculated using the expression,
$-\frac{((I+1/2)^{2} -M^2)^{1/2} \langle I+1/2||d||I-1/2
\rangle}{2(I(I+1/2)(I+1))^{1/2}}$.}
\footnotetext[3]{These shows
which Zeeman transitions have the largest E1 amplitude in a
small field. I stands for the transition $|\Psi^{(1)},-1\rangle \rightarrow
|\Psi^{(1)},0\rangle$ and II stands for the transition
$|\Psi^{(1)},0\rangle \rightarrow |\Psi^{(1)},1\rangle$. It should be noted that the difference between these amplitudes is very small.}
\footnotetext[4]{These are calculated using the expression,
$-\frac{(I+M+3/2)^{1/2}(I-M+1/2)^{1/2}I^{1/2}\langle
I+1/2||d||I-1/2\rangle}{(I+1)^{1/2}(2I+1)^{3/2}}x$, where $x = \frac{g\mu_{B}B}{\Delta E_{hf}}$.}
\footnotetext[5]{These are calculated with the expression,
$-\frac{((I+1/2)^{2}-M^2)^{1/2}\langle
I+1/2||d||I-1/2\rangle}{2(I(I+1)(2I+1))^{1/2}}$.}
\end{table*}

\begin{table*}
\caption{Here we present the M1 amplitudes as a function of field in units
of $10^{-3}e a_{B}$. We present the M1 amplitudes for the same transitions as used in Table \ref{tab:E1}.}
\begin{tabular}{c c c c }\label{tab:M1}
Atom/ion & $\langle\Psi^{(1)},M|\mu_{z}|\Psi^{(2)},M\rangle$\footnotemark[1] & $\langle\Psi,M+1|\mu_{+}|\Psi,M\rangle
$\footnotemark[2] & $\langle\Psi,M+1|\mu_{+}|\Psi,M\rangle$\footnotemark[3] \\
 & $B = 0$ & $g\mu_B B \ll \Delta E_{hf}$ & $g\mu_B B \gg \Delta E_{hf} \gg g_I \mu_N B$ \\
\hline
$^{39}$K & $ -3.2 $ & $-2.6 $ & $ -2.2/x$ \\
$^{85}$Rb & $-3.4 $ & $-2.7 $ & $ -2.4/x$  \\
$^{87}$Rb & $-3.2 $ & $-2.6 $ & $ -2.2/x$ \\
$^{133}$Cs & $-3.5$ & $-2.7$ & $-2.5/x$ \\
$^{137}$Ba$^{+}$ & $-3.2$ & $-2.6$ & $-2.2/x$ \\
$^{197}$Au & $-3.2$ & $-2.6$ & $-2.2/x$ \\
$^{205}$Tl & $-3.6$ & $-1.2$ & $-0.86/x$ \\
$^{211}$Fr & $-3.6$ & $-2.8$ & $-2.6/x$ \\
\end{tabular}
\footnotetext[1]{These are calculated using the expression $\frac{-g\mu_{B}((I+1/2)^{2}-M^{2})^{1/2}}{2I+1}$.}
\footnotetext[2]{These are calculated using the expression $\frac{-g\mu_{B}(I+M+3/2)^{1/2}(I-M+1/2)^{1/2}}{\sqrt{2}(2I+1)}$.}
\footnotetext[3]{These are calculated using the expression $-\frac{((I+1/2)^{2}-M^{2})^{1/2}g\mu_{B}}{\sqrt{2}(2I+1)x}$ where $x = \frac{g\mu_{B}B}{\Delta E_{hf}}$.}

\end{table*}

\begin{thebibliography}{20}

\bibitem{GF} J.S.M. Ginges, V.V. Flambaum. Phys. Rep. {\bf 397}, 63 (2004).

\bibitem{Flambaum} V. V. Flambaum and I. B. Khriplovich, Sov. Phys. -JEPT \textbf{52}, 835 (1980).

\bibitem{Wieman} C.S. Wood, S.C. Bennet, D. Cho, B.P. Masterson, J.L. Roberts,
C.E. Tanner, C.E. Wieman. Science {\bf 275},1759 (1997).

\bibitem{Fortson} P.A. Vetter, D.M. Meekhof, P.K. Majumder, S.K. Lamoreaux,
E.N. Fortson. Phys. Rev. Lett {\bf 74}, 2658 (1995).

\bibitem{FSK}  V. V. Flambaum, I. B. Khriplovich, and O.P. Sushkov. Phys. Lett. B{\bf 146}, 367 (1984).

\bibitem{FK} V. V. Flambaum and I. B. Khriplovich, Sov. Phys. -JEPT \textbf{62}, 872 (1985).

\bibitem{Khriplovich_1975} V. N. Novikov and I. B. Khriplovich, JEPT Lett. \textbf{22}, 74 (1975).

\bibitem{Zeeman} V.V. Flambaum, 1987 (unpublished). I.Ya. Kraftmakher. Novosibirsk Institute of Nuclear
Physics preprint INP 90-54 (unpublished).

\bibitem{Budker} D. Budker, Parity Nonconservation in Atoms, Physics Beyond the Standard Model, Proceedings of the Fifth International WEIN Symposium, P. Herczeg, C. M. Hoffman, and H. V. Klapdor-Kleingrothaus, eds. World Scientific, pp. 418-441 (1999).

\bibitem{Gorshkov} V. G. Gorshkov, V. F. Ezhov, M. G. Kozlov, and A. I. Mikahailov, Sov. J. Nucl. Phys. \textbf{48}, 867 (1988).


\bibitem{Khriplovich} I. B. Khriplovich, Parity Nonconservation in Atomic Phenomena (Gordon and Breach, New York, 1991).

\bibitem{Safronova} W. R. Johnson, M. S. Safronova, and U. I. Safronova, Phys. Rev. A \textbf{67}, 062106 (2003).

\bibitem{Happer} W. Happer, in \textit{Atomic Physics 4}, edited by G. zu Putlitz, E. W. Weber, and A. Winnacker (Plenum Press, New York, 1974), pp. 651-682.

\bibitem{Vanier} J. Vanier and C. Audoin, The Quantum Physics of Atomic Frequency Standards (Adam Hilgar, Bristol, 1989).

\bibitem{Bize} S. Bize \textit{et. al.}, Europhys. Lett. \textbf{45}, 558 (1999).

\bibitem{Blatt} R. Blatt, and G. Werth, Phys. Rev. A \textbf{25}, 1476 (1982).

\bibitem{Dahman} H. Dahman, and S. Penselin, Z. Phys. \textbf{200}, 456 (1967).

\bibitem{Lurio} A. Lurio, and A. G. Prodell, Phys. Rev. \textbf{101}, 79 (1956).

\bibitem{Lieberman} S. Liberman \textit{et. al.}, Phys. Rev. A \textbf{22}, 2732 (1980).

\bibitem{Kozlov1} M. G. Kozlov, JEPT Lett. \textbf{75}, 534 (2002).

\bibitem{Kozlov2} S. G. Porsev, and M. G. Kozlov, Phys. Rev. A \textbf{64}, 064101 (2001).

\end{thebibliography}
\end{document}